%% file: main.tex
\documentclass[conference,twocolumn]{IEEEtran}
\IEEEoverridecommandlockouts

\usepackage{ifpdf}
\usepackage{algorithmic}
\usepackage{textcomp}
\usepackage{xcolor}
\usepackage{url}
\usepackage{tabularx}
\ifpdf
  \usepackage[pdftex]{graphicx}
\else
  \usepackage[dvipdfmx]{graphicx}
\fi
\usepackage{adjustbox}
\usepackage{nameref}
\usepackage{longtable}
\usepackage{multirow}
\usepackage{threeparttable}
\usepackage{supertabular,booktabs}
\usepackage{colortbl}
\usepackage{cite}
\usepackage{amsmath}
\usepackage{subcaption}
\usepackage{footnote}
\usepackage[colorlinks=true, linkcolor=blue, citecolor=blue, urlcolor=black]{hyperref}
\def\BibTeX{{\rm B\kern-.05em{\sc i\kern-.025em b}\kern-.08em
    T\kern-.1667em\lower.7ex\hbox{E}\kern-.125emX}}
\begin{document}

\title{Implementation and Evaluation of Stable Diffusion\\ on a General-Purpose CGLA Accelerator}

\author{
    \IEEEauthorblockN{Takuto ANDO, 
    Yu ETO and
    Yasuhiko NAKASHIMA}
    \IEEEauthorblockA{Nara Institute of Science and Technology, 8916-5 Takayama-cho, Ikoma, Nara 630-0192, Japan.}
 
}

\maketitle

\thispagestyle{empty}

\input{abstract}
\input{introduction}
\input{related_work}
\input{proposed}

\input{experiments_and_results}

\input{disscution}

\input{conclusion}

\section*{Acknowledgment}

This work was supported by the JST-ALCA-Next Program (Grant Number JPMJAN23F4) and JSPS KAKENHI (Grant No. 22H00515). We also acknowledge the activities of VDEC, The University of Tokyo, in collaboration with NIHON SYNOPSYS G.K.

\bibliographystyle{IEEEtran}
\bibliography{bibliography}

\end{document}

%% file: abstract.tex
\begin{abstract}
    This paper presents the first implementation and in-depth evaluation of the primary computational kernels from the stable-diffusion.cpp image generation framework on IMAX3, a general-purpose Coarse-Grained Reconfigurable Array~(CGRA) accelerator. 
    We designed IMAX3 as a versatile computational platform, and this work assesses its capabilities by executing a demanding image generation workload. 
    We evaluate its performance on a current Field-Programmable Gate Array~(FPGA) prototype to establish a baseline and project its potential for a future Application-Specific Integrated Circuit~(ASIC) implementation. 
    Our results demonstrate that, despite its general-purpose architecture, IMAX3 achieves promising performance and power efficiency, particularly in its projected ASIC form. 
    This work provides concrete guidelines for future IMAX architectural designs and establishes a foundation for developing next-generation, AI-specialized Coarse-Grained Linear Array~(CGLA) accelerators by refining this versatile platform. 
    Ultimately, this achievement contributes to the realization of energy-efficient, on-device, multi-modal AI platforms.

    \end{abstract}
    
    \begin{IEEEkeywords}
    Stable Diffusion, Image generation, FPGA, CGRA, CGLA, IMAX
    \end{IEEEkeywords}
    

%% file: introduction.tex
\section{Introduction}
In recent years, generative Artificial Intelligence~(AI) models, particularly diffusion-based models, have driven innovation across diverse fields by generating high-quality content\cite{Xiao_2025_CVPR,Hu_2024_CVPR,llm4gen,diffit,NEURIPS2023_821655c7}. 
However, the sophistication of these models demands vast computational resources. 
Specifically, the requirements for high computational throughput and wideband memory access have led to significant increases in power consumption.

Currently, the execution of Large Language Models~(LLMs) and generative AI models heavily relies on General-Purpose Graphics Processing Units~(GPGPUs) due to their high parallel processing capabilities. 
However, their widespread adoption has significantly increased the energy footprint of data centers. 
A report by Shehabi et al. suggests that by 2028, data centers in the United States could account for $6.7\,\mathrm{\%}$ to $12\,\mathrm{\%}$ of the nation's projected electricity consumption\nobreak\cite{shehabi2024united}. 
As conventional CPU and GPU architectures approach their limits in improving power efficiency, the development of novel, energy-efficient computer architectures is critical.

\begin{figure}[t]
  \centering
  \includegraphics[width=0.98\columnwidth]{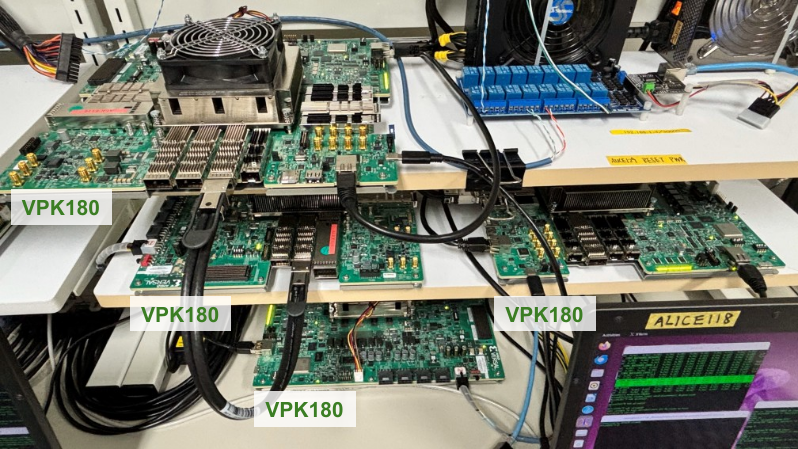}
  \caption{ The IMAX3 FPGA prototype, featuring a multi-board configuration with four AMD Versal VPK180 evaluation kits.}
  \label{fig:imax3_proto}
  \vspace{-1em}

\end{figure}

To address these challenges, Akabe et al. proposed In-Memory Accelerator eXtension 3~(IMAX3)\nobreak\cite{imax_access}, an accelerator architecture based on a Coarse-Grained Reconfigurable Array~(CGRA) that aims to achieve both high power efficiency and programmatic flexibility. 
Fig.~\ref{fig:imax3_proto} shows the IMAX3 prototype.
IMAX3 utilizes a Coarse-Grained Linear Array~(CGLA) architecture, an evolution of the CGRA, which connects processing units in a linear structure. 
This architecture strategically interleaves processing units with cache memory to absorb irregular memory access latencies and maximize the utilization of Local Memory Modules~(LMM). 
Furthermore, the design of IMAX3 focuses on enabling flexible placement of functional units and employing logically aligned execution patterns to achieve efficient computation. 
Therefore, IMAX3 is not specialized for specific AI tasks, but is designed as a general-purpose computing accelerator platform capable of executing a variety of computing kernels.

Previous work demonstrated the effectiveness of IMAX3 for various AI workloads, such as CNNs\nobreak\cite{unetimax,imaxcnn2,imaxcnn3} and LLMs\nobreak\cite{uetanicgra,eto2025implementation}. 
Building on this foundation, we extend its application to the domain of image generation. 
We extract the core quantized dot-product kernels from the stable-diffusion.cpp framework, implement them on the general-purpose IMAX3 accelerator, and evaluate performance. 
This work has two main objectives.
First, we execute the core dot-product operations of image generation on the IMAX3 architecture to clarify its real-world performance on a current Field-Programmable Gate Array~(FPGA) prototype. 
Second, we analyze the performance potential achievable with a future Application-Specific Integrated Circuit~(ASIC) implementation.

\begin{figure*}[t]
  \centering
  \includegraphics[width=1.9\columnwidth]{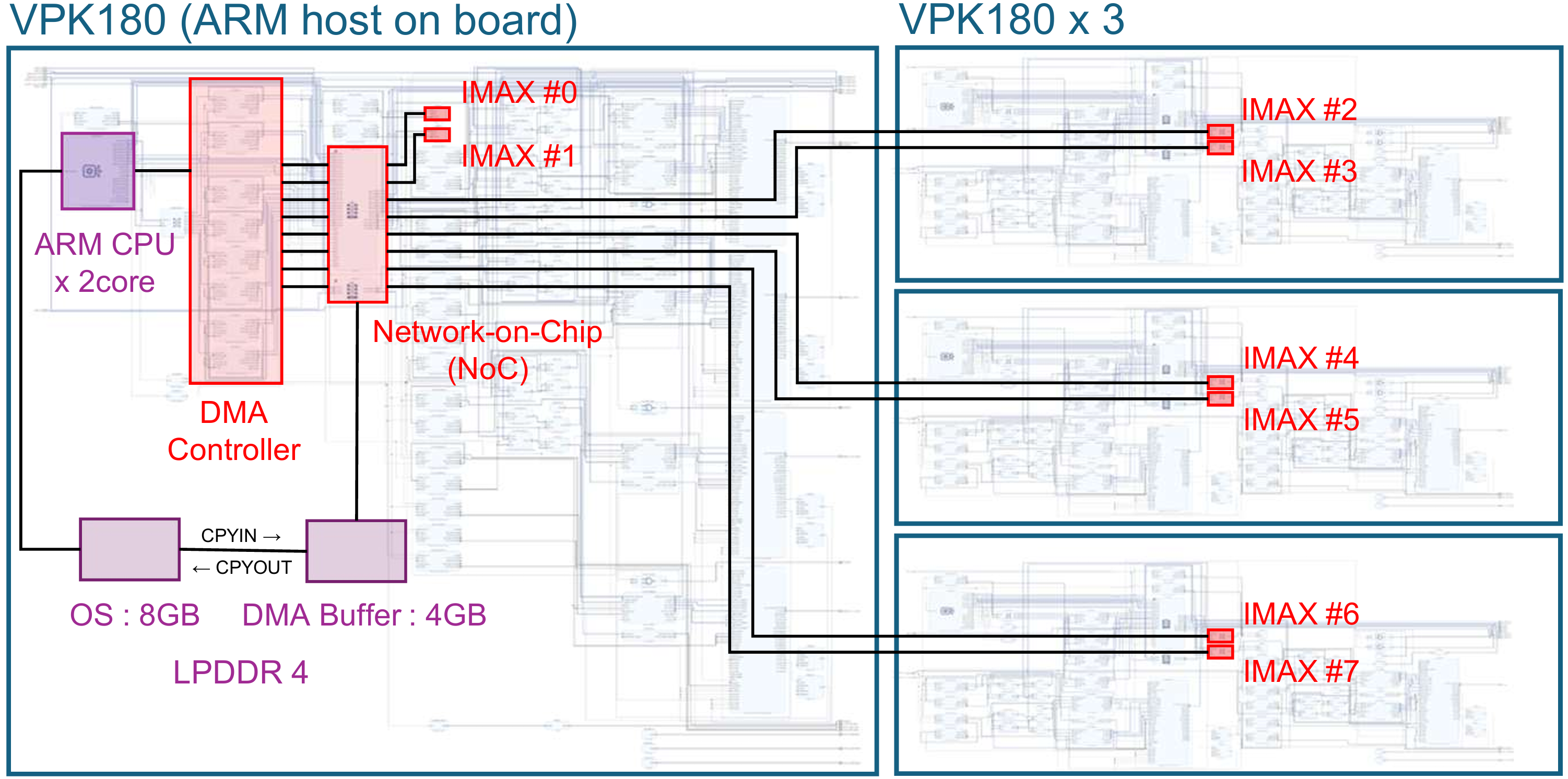}
  \caption{High-level overview of the IMAX3 system architecture, implemented on a multi-FPGA platform with four AMD Versal VPK180 devices.}
  \label{fig:imax3_conf}

\end{figure*}

This work not only demonstrates the versatility of IMAX but also establishes a crucial foundation for developing future AI-specialized CGLA accelerators. 
The insights we gain will enable us to design dedicated, high-performance accelerators by starting from the IMAX platform, streamlining its instruction set to remove redundant features, and enhancing its AI-specific capabilities.

The main contributions of this paper are as follows:
\begin{itemize}
 \item To the best of our knowledge, we are the first to implement and evaluate the primary computational kernels of Stable Diffusion on a general-purpose CGRA, providing a crucial performance baseline for this research domain.
 \item By successfully implementing kernels for both language and image generation models, we demonstrate that the IMAX architecture possesses the versatility to handle diverse AI domains, demonstrating its potential as a multi-modal AI platform.
 \item Through performance projections for an ASIC implementation, we show that IMAX and its derivative CGLA accelerators can become viable, energy-efficient alternatives to GPUs, particularly for power-constrained AI applications.
\end{itemize}

The rest of this paper is organized as follows.
Section~\ref{rwork} provides related work on AI accelerators and describes the previous work on the IMAX3 architecture.
Section~\ref{proposed} describes the implementation method of Stable Diffusion on IMAX3
Section~\ref{ex_and_re} presents the results of our performance evaluations, including fundamental characterization via image generation using Stable Diffusion.
Section~\ref{discussion} provides a detailed analysis of the dot-product execution on IMAX3. 
Finally, Section~\ref{conclusions} concludes our work.

%% file: related_work.tex
\section{Related Work}
\label{rwork}

\subsection{Hardware Accelerators for AI Image Generation}
In both the training and inference of generative AI, GPUs are widely utilized due to their high parallel computing capabilities. 
However, deploying GPUs entails significant initial costs and substantial power consumption, presenting a major challenge, particularly for systems dedicated to specific inference tasks. 
This background has led to a strong demand for the development of alternative hardware architectures that achieve higher power efficiency.

To address this challenge, research into hardware accelerators using reconfigurable devices, such as FPGAs\nobreak\cite{fpgacnn,vitisai,multi_task} and CGRA\nobreak\cite{cgrasurvey,cgra_cnn2,hmap,cgra_crypto}, is actively progressing. 
These are dedicated accelerators optimized for specific models and operations, delivering high performance for target tasks.
In the domain of LLMs, for instance, several FPGA-based accelerators have been proposed\nobreak\cite{llm4fpga,llamaf,zeng2024flightllm,llama2fpga}. 
Similarly, for image generation, SD-Acc\nobreak\cite{sdacc} and SDA\nobreak\cite{sda} have been reported as an FPGA-based accelerator specifically for Stable Diffusion. 
In contrast to these specialized accelerators, the IMAX3 platform used in this work is a general-purpose accelerator designed to handle a diverse range of operations, not limited to a specific AI model or kernel. 
This flexibility offers significant advantages in rapidly evolving AI algorithms and the implementation of multiple different AI tasks. 
\subsection{IMAX Architecture}

CGRAs offer a promising architectural paradigm, aiming to combine the efficiency of ASICs with the flexibility of CPUs and GPUs. 
However, conventional CGRAs have faced persistent challenges, including scalability limitations, long compilation times, and the complexity of implementing loops\nobreak\cite{cgra1}. 
To address these limitations, our previous work proposed IMAX3, an accelerator based on a CGLA, which is an evolution of the CGRA\nobreak\cite{imax_access}.

The fundamental design of the IMAX architecture strategically interleaves Processing Elements~(PEs) and cache memory in a linear array structure. 
This arrangement enables the architecture to absorb irregular memory access latencies and maximize the utilization of LMM. 
In addition, this design mitigates memory access bottlenecks\nobreak\cite{imax_access}. 
Each PE contains a pipelined structure, comprising an Arithmetic Logic Unit~(ALU) and local memory, to sustain high throughput.

As shown in Fig.~\ref{fig:imax3_conf}, IMAX3 implements this 8-lane configuration on an AMD Versal VPK180x4.
At the system level, we implement IMAX3 as a System-on-Chip~(SoC) where a Processing System~(PS) and Programmable Logic~(PL) connect via a Network-on-Chip. 
The PS, featuring an ARM Cortex-A72 CPU, manages the operating system and overall system control, while the PL hosts the CGLA core comprising multiple PEs.
Furthermore, IMAX3 features a multi-lane configuration. 
This design achieves high processing throughput by assigning lanes according to an application's degree of parallelism, enabling parallel task execution. 
The IMAX design philosophy of logically aligned execution patterns contributes to achieving a balance between high compilation efficiency and high execution efficiency.

In previous works, we established IMAX3 as a general-purpose computational platform, demonstrating its effectiveness across a range of workloads. 
These include not only traditional computational kernels like SpGEMM and FFT\nobreak\cite{imax_access} but also modern AI workloads such as CNNs\nobreak\cite{unetimax,imaxcnn2,imaxcnn3} and LLMs\nobreak\cite{uetanicgra,eto2025implementation}.
Building on these results, this work extends the application domain of IMAX3 to image generation. 
Specifically, we execute the dot-product operations, which form the computational core of this task, on the IMAX3 architecture.

%% file: proposed.tex
\section{Implementation of Stable Diffusion on IMAX}
\label{proposed}
In this section, we describe the implementation of Stable Diffusion on IMAX.
  
\subsection{stable-diffusion.cpp}

Stable Diffusion\nobreak\cite{rombach2022highresolutionimagesynthesislatent} is an image generation method based on diffusion models. 
It works by learning to restore original data from noisy data and then uses the reverse of that process to generate images from random noise.
While Stable Diffusion is widely adopted for its ability to generate high-quality content, its iterative denoising process requires significant computational resources.

In this work, we select stable-diffusion.cpp\cite{stable-diffusion.cpp} as the performance evaluation framework. 
This is a C/C++ implementation of the Stable Diffusion v1.5 model based on the GGML tensor library.
We chose GGML as it is a tensor library designed for minimal external dependencies, low-level hardware optimization, and efficient memory management. 
It enables flexible deployment across diverse hardware platforms, including CPUs and GPUs, and excels at high-performance matrix operations and quantization. 
Using stable-diffusion.cpp allows us to eliminate the complex software overhead associated with high-level frameworks like PyTorch. 
This approach allows for a more direct and accurate evaluation of the IMAX accelerator's performance.

Adopting the GGML framework is a strategic decision for the IMAX architecture. 
Our previous work implementing llama.cpp\cite{GerganovLlamaCpp2023} on IMAX3 also leveraged this same tensor library. 
This common foundation creates a significant advantage because we can directly reuse the quantized dot-product kernels developed for LLM inference in the current image generation workload.
This capability demonstrates that IMAX can flexibly address AI tasks from different domains, such as language and vision, using a single hardware design and a shared library of compiled kernels. 
Consequently, this approach significantly reduces the development cost for new applications and further demonstrates IMAX's capability as a true general-purpose accelerator.

\begin{table}[t]
  \centering
  \caption{Breakdown of execution time in dot-product kernel}
  \label{tab:kernel_performance}
  \setlength{\tabcolsep}{10pt} 
  \renewcommand{\arraystretch}{1.2} 
  \begin{tabular}{rcccc} 
      \hline\hline
      \textbf{Quantized Type} & \textbf{F32} & \textbf{F16} & \textbf{Q3\_K} & \textbf{Q8\_0} \\
      \hline
      Q3\_K Model  & 30.7\,\% & 59.0\,\% & 10.3\,\% & -- \\
      Q8\_0 Model    & 21.8\,\% & 62.0\,\% & -- & 16.3\,\% \\
      \hline
  \end{tabular}
  \vspace{-1em}
\end{table}
\subsection{Offloading of Quantized Dot-Product}
This subsection describes our method for offloading the quantized dot-product operations of Stable Diffusion to the IMAX.
The denoising process within the U-Net component consumes the majority of the computational effort in Stable Diffusion. 
The U-Net heavily utilizes convolutional layers and attention mechanisms, and the dot-product operation forms their fundamental arithmetic core.
In this work, we target this computationally intensive dot-product for offloading to IMAX. 
Specifically, we reuse the quantized dot-product kernels previously developed for the GGML framework in our previous work on LLMs\cite{eto2025implementation}. 
We implement two distinct quantization schemes, each offering a different trade-off between model size and precision: Q8\_0 (8-bit integer quantization) and Q3\_K (3-bit k-quants method). 
The GGML framework widely employs these kernels to reduce memory footprint and accelerate computation on target hardware.

However, our execution profiling of stable-diffusion.cpp reveals that the framework also frequently calls non-quantized dot-product operations in FP16 and FP32 formats. 
Table \ref{tab:kernel_performance} shows the proportion of total dot-product execution time attributable to each data type, calculated from pure computation time with memory copy overhead excluded. 
As the data indicates, the computational share of quantized dot-product is smaller than that of the F32 and F16 counterparts.
In our current implementation, we do not offload these non-quantized kernels. 
Instead, they execute on the host CPU.
Therefore, our offloading strategy addresses only a portion of the model's overall computational bottleneck. 
The implementation of FP16 and FP32 kernels remains a critical task for future work.

Despite the currently limited offload ratio, our strategic focus on quantized kernels is forward-looking. 
For deploying large-scale generative AI models like Stable Diffusion on resource-constrained edge devices, model quantization is not merely an optimization but a necessity. 
As on-device AI becomes more prevalent, low-bit integer kernels, such as the ones we implemented, are expected to become the mainstream computational workload.

\begin{figure}[t]
  \centering
  \includegraphics[width=1.0\columnwidth]{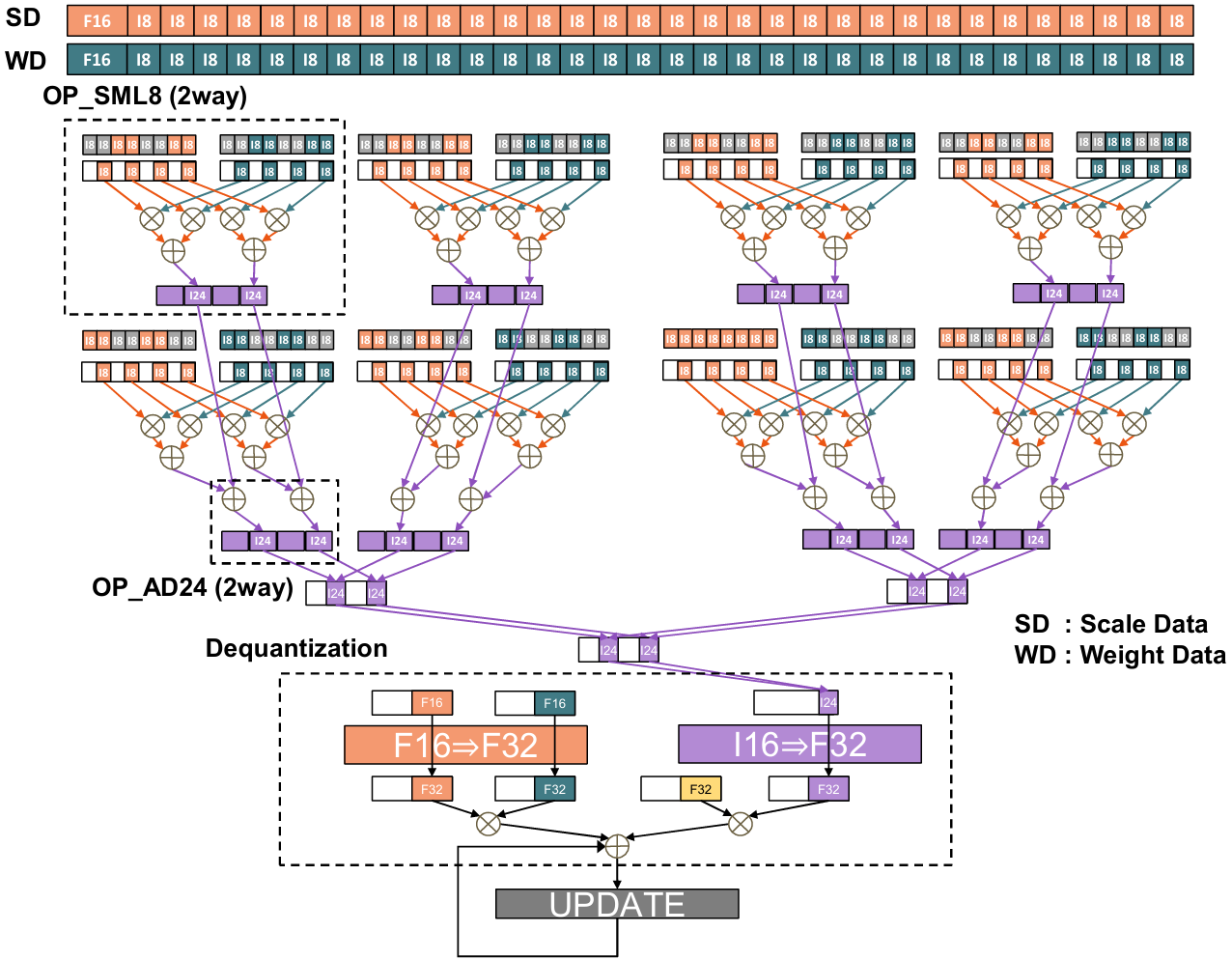}
  \caption{Processing flow of Q8\_0 kernel.}
  \label{fig:q8dot}
  
  \end{figure}

\begin{figure}[t]
  \centering
  \includegraphics[width=1.0\columnwidth]{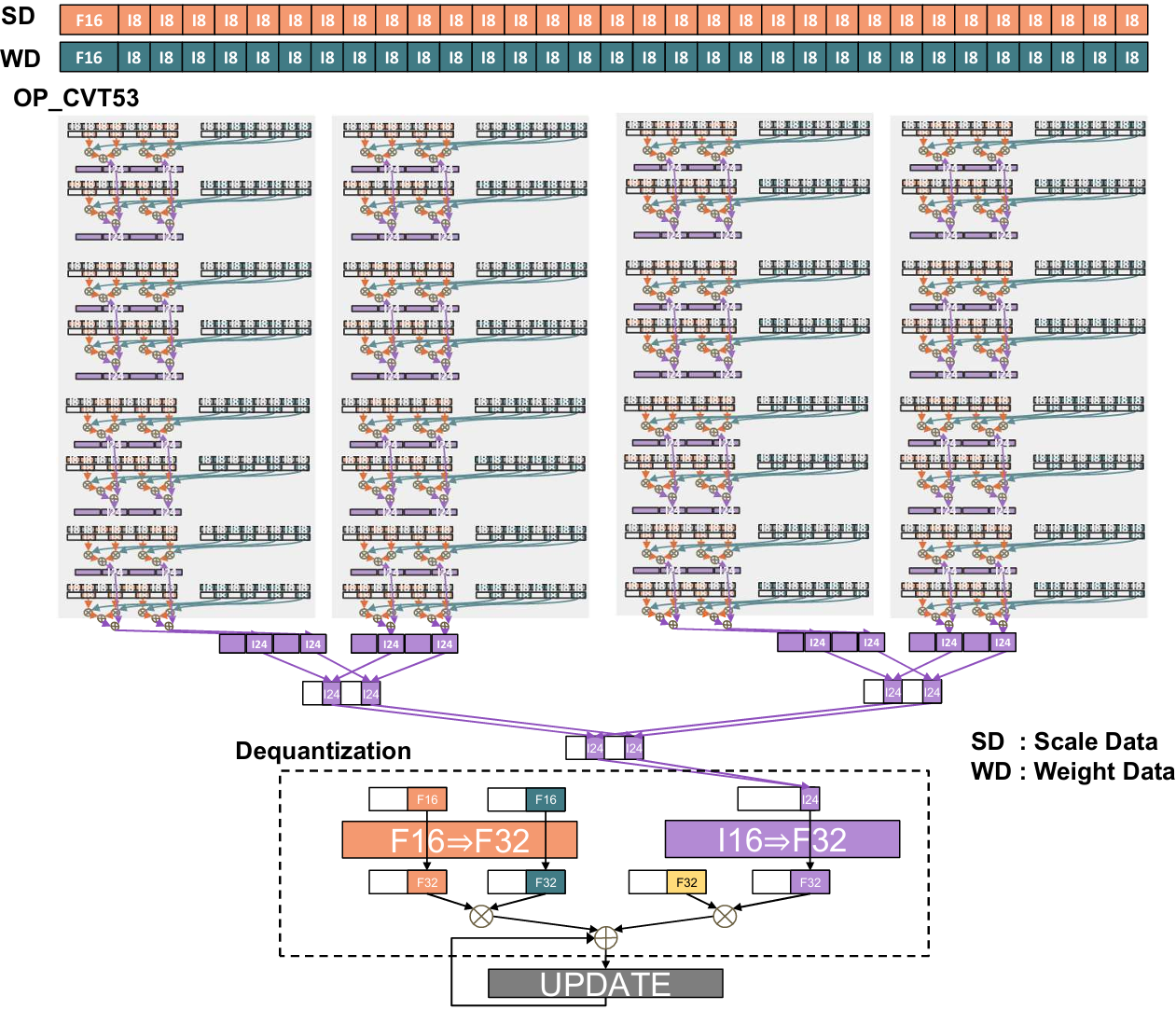}
  \caption{Processing flow of Q3\_K kernel.}
  \label{fig:q3kdot}
  
  \end{figure}

We optimized the data flow of the quantized dot-product kernels for the IMAX pipeline architecture, causing it to differ from the original GGML implementation. 
As shown in Fig.~\ref{fig:q8dot} and Fig.~\ref{fig:q3kdot}, the IMAX dataflow efficiently combines multiplication, addition, and data type conversion. 
For the Q8\_0 kernel, the process aggregates the results of 8-bit integer multiplication and addition into a 24-bit integer across every 12\,PEs. 
It then performs a final multiplication with a single-precision 32-bit floating-point value.

In contrast, implementing the Q3\_K kernel requires more complex data manipulation. 
The Q3\_K quantization format natively uses 6-bit scale data alongside 2-bit and 1-bit quantized data. 
To process this data efficiently on IMAX's SIMD architecture, we designed a custom data restructuring method. 
Specifically, we convert the 6-bit scale data to 5-bit and pack the 2-bit and 1-bit segments into a unified 3-bit format. 
This restructuring allows us to combine the 8-bit input data with 5-bit and 3-bit weight data, creating an operational flow similar to that of the Q8\_0 kernel. 
In addition, we have empirically confirmed that approximating scale data has almost no effect on the final calculation results.

  \begin{figure}[t]
    \centering
    \begin{subfigure}[b]{0.4\columnwidth} 
        \centering
        \includegraphics[width=\textwidth]{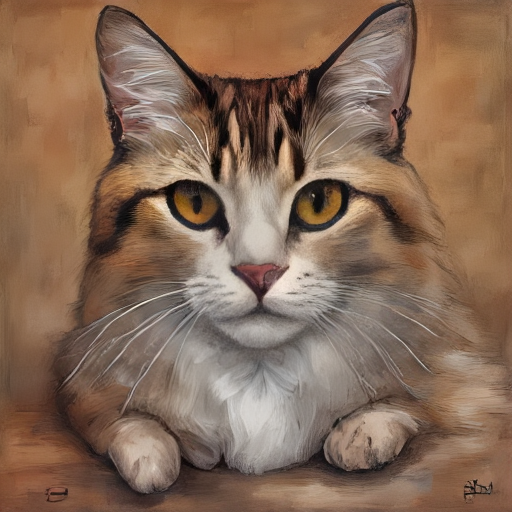}
        \caption{Q3\_K model}
        \label{fig:subfigA}
    \end{subfigure}
    \hspace{1em} 
    \begin{subfigure}[b]{0.4\columnwidth} 
        \centering
        \includegraphics[width=\textwidth]{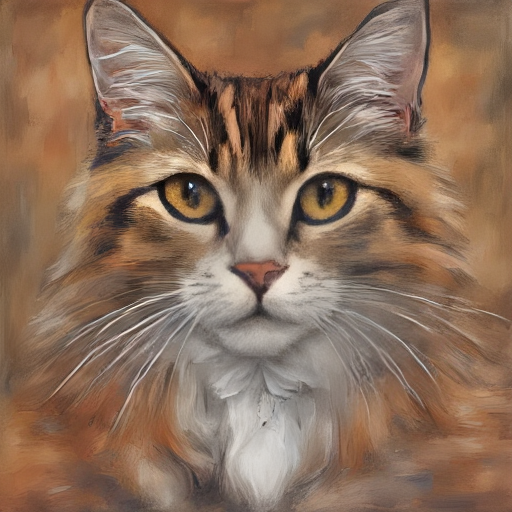}
        \caption{Q8\_0 model}
        \label{fig:subfigB}
    \end{subfigure}
    \caption{Generated images of Q3\_K and Q8\_0 models.}
    \label{fig:sd_turbo_example}
    \vspace{-1em}
  \end{figure}

To realize these optimized processing flows, the following dedicated instructions have been added to the IMAX instruction set:
\begin{itemize}
\item \textbf{OP\_SML8}: A 2-way SIMD signed 8-bit integer multiply-add instruction. It independently multiplies the 8-bit segments of its input operands and sums the results, producing a sign-extended 24-bit output.
\item \textbf{OP\_AD24}: A 2-way 24-bit integer addition instruction, which we use to aggregate the intermediate results from OP\_SML8.
\item \textbf{OP\_CVT53}: A SIMD conversion instruction specialized for the Q3\_K quantization scheme. It performs the aforementioned data restructuring, converting inputs of different bit widths (5-bit and 3-bit) into a single format, and executes scaling and signed multiplication in parallel.
\end{itemize}
These custom instructions enable efficient, hardware-level processing of GGML's complex quantized data structures and maximize functional unit utilization. 
We map the Q3\_K kernel across 51 of the 64 PEs and the Q8\_0 kernel across 46 PEs. 
This mapping strategy allows IMAX to achieve high performance on these specific kernels while maintaining its core general-purpose flexibility.


    

  


\begin{table*}[t]
  \centering
  \begin{threeparttable}
  
  \caption{Physical specifications and performance comparisons of evaluated hardware platforms.}
  \label{tab:processor_comparison_annotated}
  
  \begin{tabular}{@{} llrrrrrrr l @{}} 
    \hline\hline
    \textbf{Device} & \textbf{CPU} & \textbf{Cores} & \textbf{Chip area} & \textbf{Process} & \textbf{Operating frequency} & \textbf{Memory} & \textbf{Power}\tnote{d} \\
    & & & \textbf{(mm$^2$)} & \textbf{node} & & & \textbf{(W)} \\
    \hline
 
    \textbf{ARM Cortex-A72 (on Versal)} & - & 2 & - & 7\,nm & 1.4\,GHz & 8\,GB DDR4 & 1.5 \\
 
    \textbf{IMAX3 (Xilinx VPK180)} & ARM Cortex-A72 & 64\tnote{a} & - & 7\,nm & 145\,MHz & 8\, + 4\,GB DDR4\tnote{c} & 180 \\
 
    \textbf{IMAX3 (28nm)} & - & 64\tnote{a} & 14.6 & 28\,nm & 800\,MHz& - & 47.7 or 52.8 \\
 
    \textbf{Intel Xeon w5-2465X} & - & 16 & - & Intel 7 & 3.1\,GHz& 512\,GB DDR5 & 200 \\
 
    \textbf{NVIDIA GTX 1080 Ti} & Xeon w5-2465X & 3584\tnote{b}& 471 & 16\,nm & 1480\,MHz & 11\,GB GDDR5X & 250 \\
    \hline
  \end{tabular}
  

 \begin{tablenotes}[para,flushleft]
   \item[a] The number of cores for IMAX3 refers to the number of PEs per lane.

   \item[b] The number of cores for the NVIDIA GTX 1080 Ti refers to the number of CUDA cores.

   \item[c] 8\,GB DDR4 for OS buffer and 4\,GB DDR4 for DMA buffer.

   \item[d] IMAX3~(28\,nm) is an estimated value, references for other devices are from Cortex-A72\nobreak\cite{Humrick_CortexA72_2016}, Intel Xeon\nobreak\cite{Intel_Xeon_w5-2465X_Specs}, NVIDIA GTX 1080 Ti\nobreak\cite{NVIDIA_Turing_2018}.
 \end{tablenotes}
  
   \end{threeparttable}
  \end{table*}

%% file: experiments_and_results.tex
\section{Experiments and Results}
\label{ex_and_re}
In this section, we evaluate the performance of the primary computation kernels of stable-diffusion.cpp executing on the IMAX3 accelerator. 
There are two objectives of this experiment.
First, we clarify the real-world performance of our current FPGA prototype. Second, we leverage this empirical data along with logic synthesis results to project the performance potential of a future ASIC implementation.

\begin{figure}[t]
  \centering
  \includegraphics[width=1.0\columnwidth]{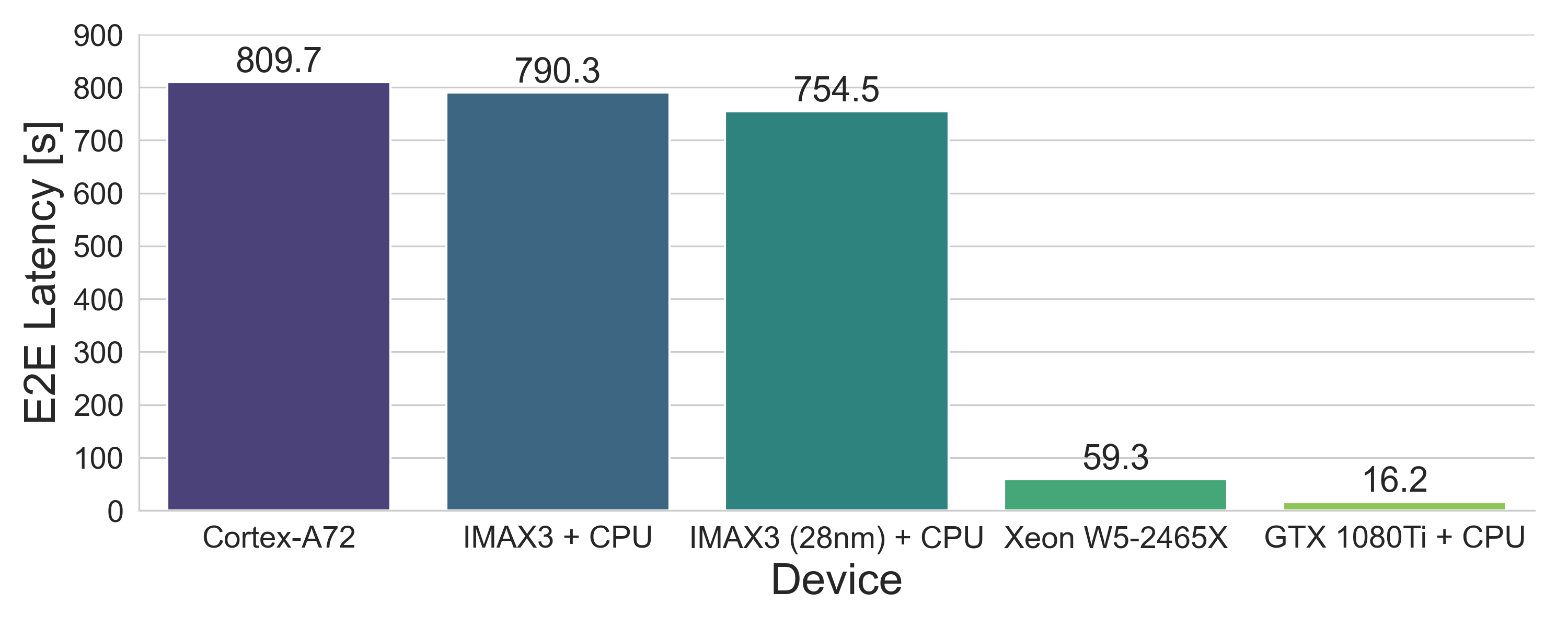}
  \caption{End-to-end latency comparison by device\\ for Q3\_K model inference.}
  \label{fig:performance_plot_q3ks}

\end{figure}
\begin{figure}[t]
  \centering
  \includegraphics[width=1.0\columnwidth]{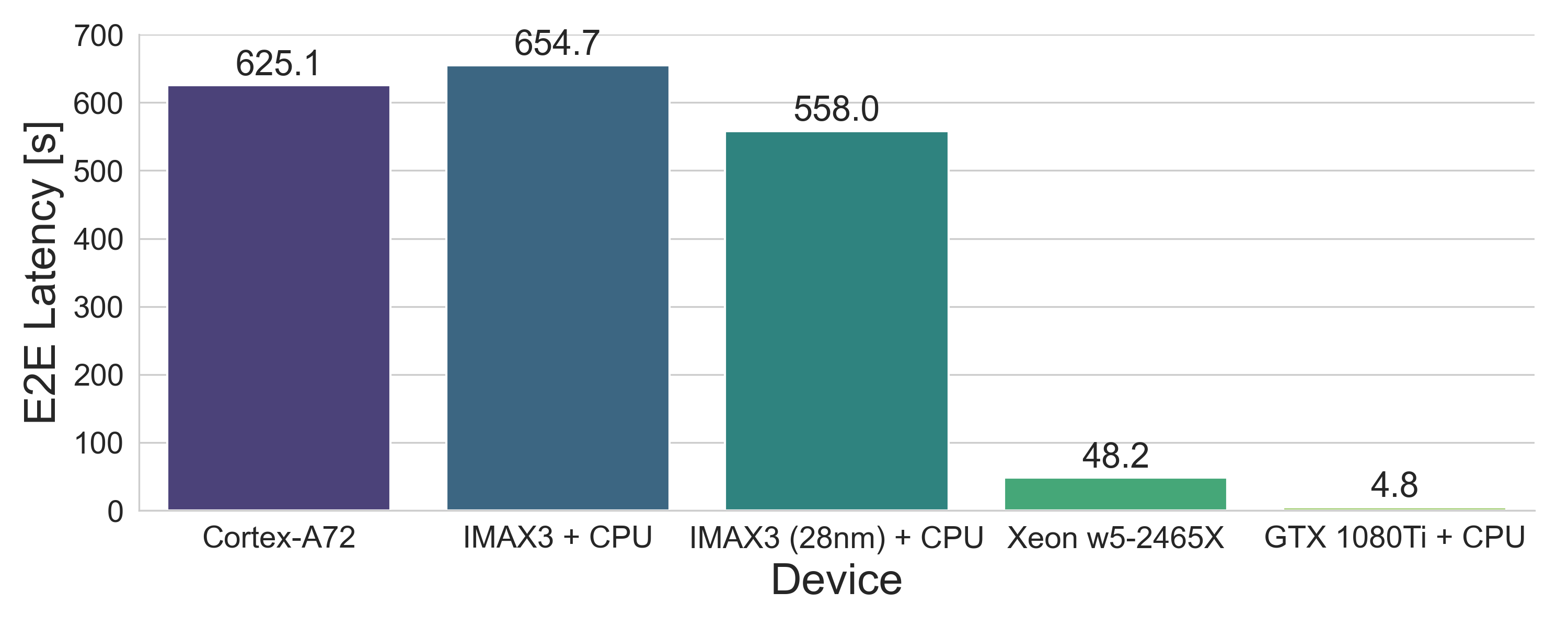}
  \caption{End-to-end latency comparison by device\\ for Q8\_0 model inference.}
  \label{fig:performance_plot_q8}

\end{figure}

\begin{figure}[t]
  \centering

  \includegraphics[width=1.0\columnwidth]{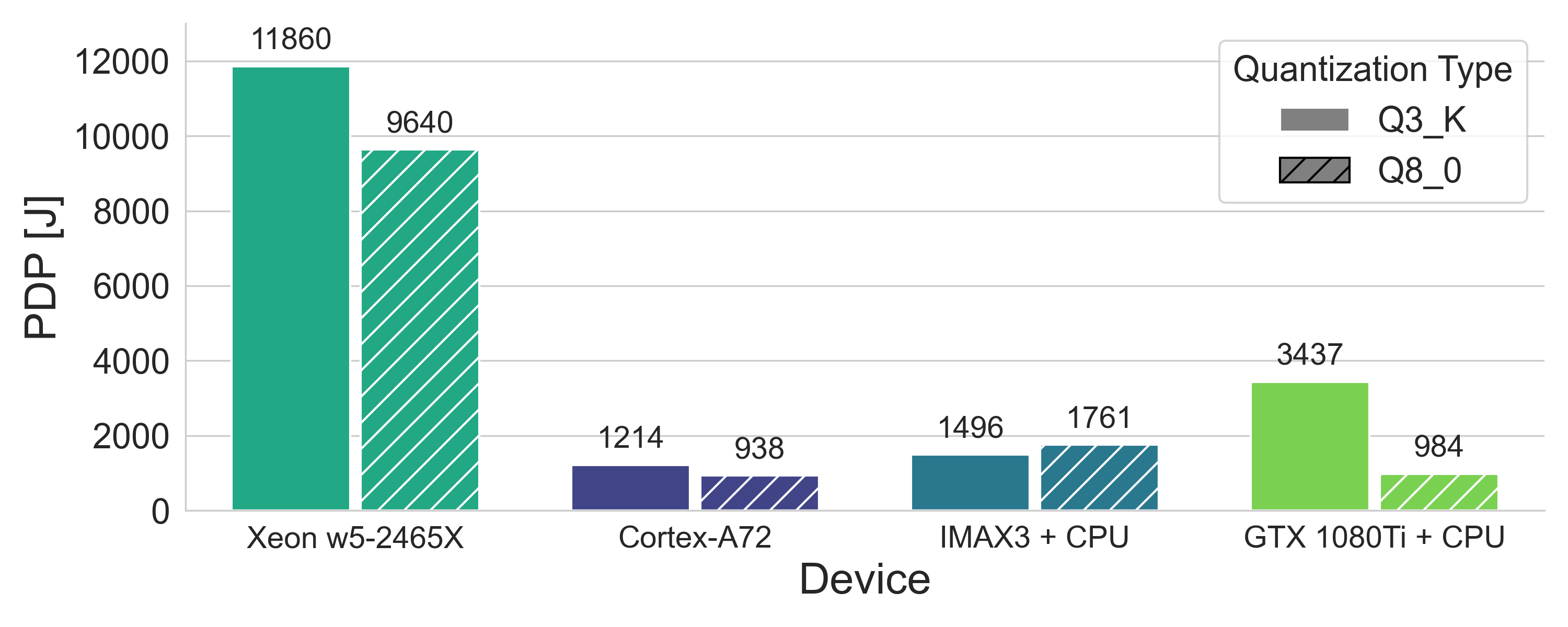}

  \caption{PDP comparison with each device.}
  \label{fig:microbenchmark_time}

\end{figure}
\subsection{Experimental Setup}

We measured performance by generating a 512$\times$512 pixel image using the stable-diffusion.cpp framework. 
The experiment used the SD-Turbo model\cite{sauer2023adversarialdiffusiondistillation}, a single inference step, and the prompt "a lovely cat." 
The SD-Turbo model can produce high-quality images in a single step.
Fig.~\ref{fig:sd_turbo_example} shows an example of the generated output.

We conducted our hardware evaluation on a prototype implemented on an AMD Versal Premium VPK180 evaluation kit. 
We developed the SoC design using the Vivado 2024.1 tool, with the IMAX core described in Verilog HDL. 
This SoC integrates a PS featuring a dual-core ARM Cortex-A72 CPU with PL that hosts the 64\,PE IMAX computational unit. 
In this evaluation, IMAX implemented on PL was used in a single-lane configuration at an operating frequency of 145 MHz.

For comparison, we benchmarked against the host ARM Cortex-A72, a high-performance Intel Xeon w5-2465X CPU, and an NVIDIA GeForce GTX 1080 Ti GPGPU. 
Table~\ref{tab:processor_comparison_annotated} summarizes the key characteristics of each device, showing that IMAX3 has significantly fewer processing cores (64\,PEs per lane) compared to the GPU's thousands of cores (3,584\,cores). 
To calculate power consumption, we used the nominal Thermal Design Power~(TDP) values for the Intel Xeon w5-2465X\nobreak\cite{Intel_Xeon_w5-2465X_Specs} and NVIDIA GTX 1080 Ti\nobreak\cite{NVIDIA_Turing_2018}, and an estimated value for the ARM Cortex-A72\nobreak\cite{Humrick_CortexA72_2016}. 
We estimated the power consumption of IMAX from the results of logic synthesis using Synopsys Design Compiler with a TSMC 28\,nm process library. 
The power draw of IMAX depends on the LMM size and the number of active functional units. 
In our 512\,KB LMM configuration, we estimated the power at 47.7\,W for the Q8\_0 kernel~(46~units) and 52.8\,W for the Q3\_K kernel~(51~units).
To project the performance of an ASIC implementation, we conducted a static timing analysis using our 28\,nm synthesis results. 
The analysis, performed with Synopsys Design Compiler\nobreak\cite{SynopsysNDDCUltra}, revealed that the critical path delay allows for a maximum operating frequency of 840 MHz, which we adopted for our projection. 
This frequency is therefore not an assumption but a determination based on a physical design exploration. 
Based on this 840\,MHz clock frequency, we project an approximate 5.8x reduction in IMAX's computation time compared to the 145\,MHz FPGA version.

\begin{figure*}[t]
  \centering

  \includegraphics[width=2\columnwidth]{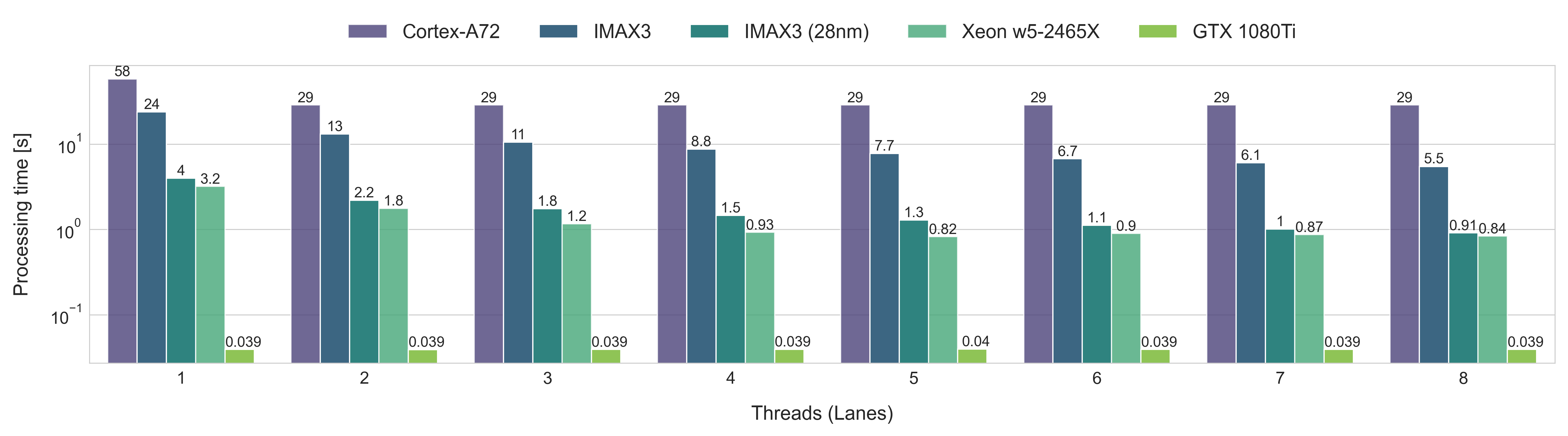}
  \caption{Comparison of execution times for each device in the Q3\_K kernel.}
  \label{fig:compare_kernel_q3k}

\end{figure*}

\begin{figure*}[t]
  \centering

  \includegraphics[width=2\columnwidth]{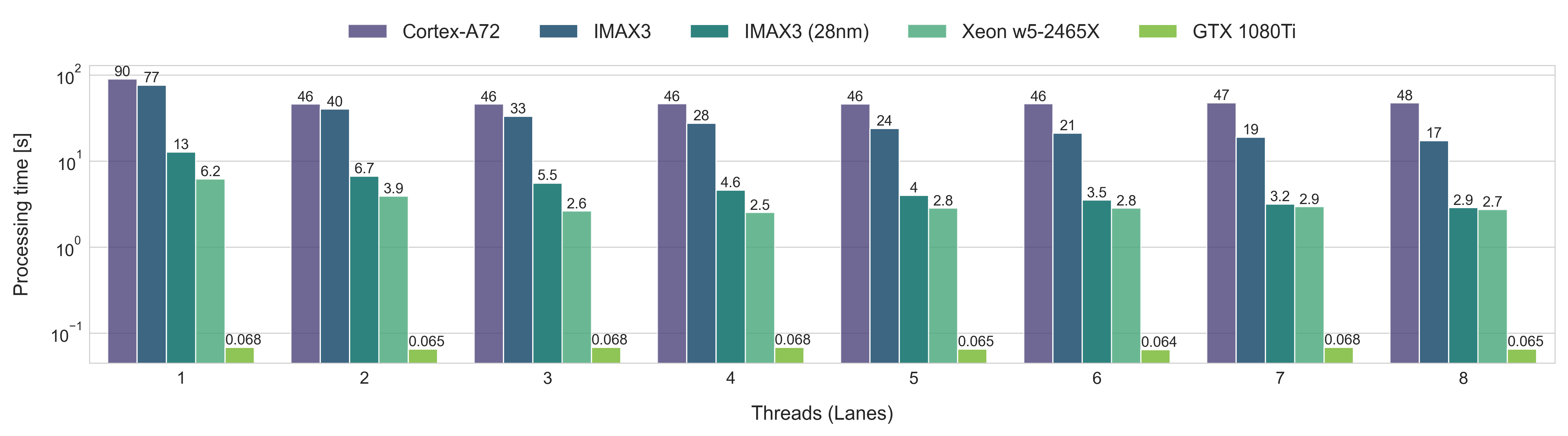}
  \caption{Comparison of execution times for each device in the Q8\_0 kernel.}
  \label{fig:compare_kernel_q80}

\end{figure*}


\begin{figure*}[t]
  \centering

  \includegraphics[width=2\columnwidth]{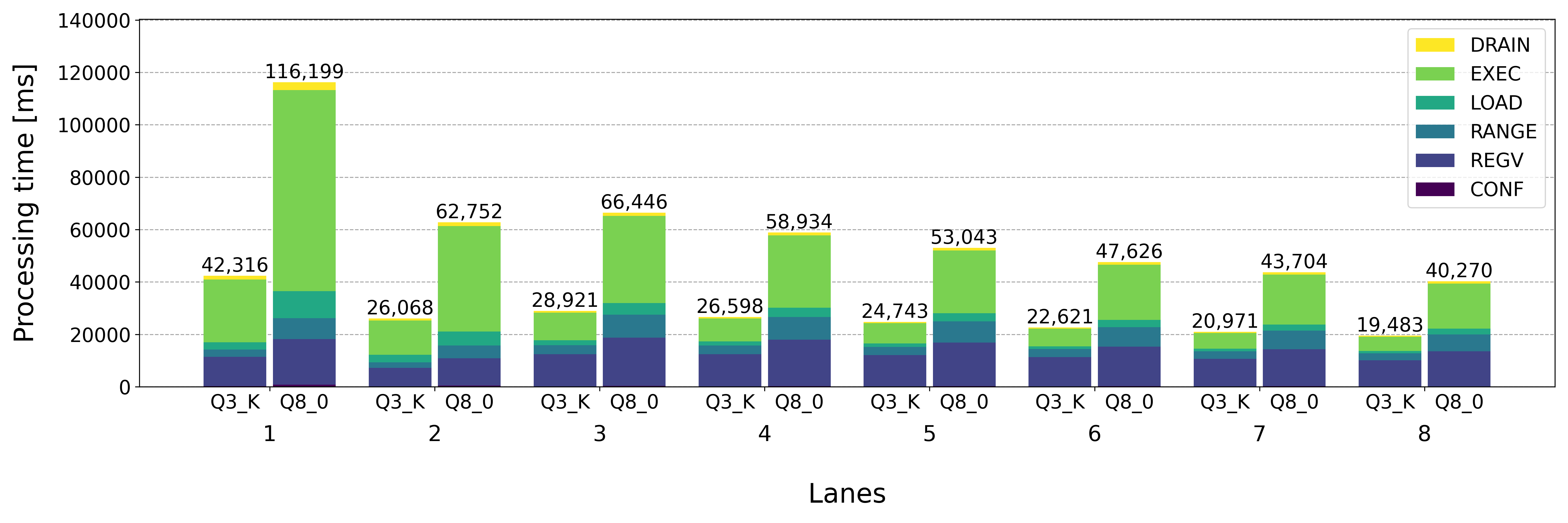}

  \footnotesize
  \begin{tabular}{@{} l l @{\hspace{1cm}} l l @{\hspace{1cm}} l l @{}}
      \textbf{CPU}:   & CPU processing                    & \textbf{DRAIN}: & Cache $\Rightarrow$ Main memory transfer & \textbf{CONF}: & IMAX command transfer  \\
      \textbf{REGV}:  & Register initialization          & \textbf{LOAD}: & Main memory $\Rightarrow$ Cache transfer& \textbf{EXEC}:  & Burst calculation     \\
  \end{tabular}

  \caption{IMAX processing time breakdown comparison of Q3\_K and Q8\_0 kernels on the FPGA.}
  \label{fig:microbenchmark_time}

\end{figure*}


\subsection{Evaluation of Processing Time and Power Consumption}
Fig.~\ref{fig:performance_plot_q3ks} and Fig.~\ref{fig:performance_plot_q8} show the end-to-end~(E2E) latency for image generation using both the Q3\_K and Q8\_0 models.
When using the Q3\_K model, the FPGA-based IMAX achieved a latency of 790.3\,s, outperforming the 809.7\,s required for standalone execution on the host ARM CPU. 
However, even with a projection for an 840\,MHz ASIC, the estimated latency of 754.5\,s remains significantly longer than that of the Xeon CPU at 59.3\,s and the GPU at 16.2\,s. 
This performance gap is primarily due to a limited offload ratio of less than 20\% referring to Table~\ref{tab:kernel_performance}. 
Consequently, the host ARM CPU's execution time continues to dominate the overall end-to-end latency.
For Q8\_0 quantization, the larger data transfer volume degraded the FPGA version's performance to 654.7\,s, falling behind the 625.1\,s of the standalone ARM execution. 
The projected ASIC implementation, however, overcomes this by substantially accelerating the offloaded portion, which reduces the total latency to 558.0\,s. 
Despite this gain, the overall latency for the complete process still does not approach the performance levels of the Xeon CPU or the GPU.

Next, we evaluate the energy efficiency of each hardware platform benchmarked. 
We adopt the Power-Delay Product~(PDP) as our metric, which considers both execution time and power consumption. 
The PDP is defined by the following equation (\ref{eq:pdp}), where a lower value signifies superior energy efficiency, indicating the ability to complete a task faster with less total energy.
\begin{equation}
\label{eq:pdp}
PDP = \text{Execution Time} \times \text{Power}
\end{equation}
In our evaluation, we calculated the PDP by considering the power consumption during each distinct execution phase for different devices, reflecting the total energy consumption of the system.
The evaluation results show that the low-power ARM Cortex-A72 exhibited the lowest PDP. 
Notably, the projected PDP for the ASIC version of IMAX significantly surpassed that of the high-performance Xeon CPU for both the Q8\_0 and Q3\_K models. 
In the Q3\_K case, IMAX~(28\,nm) achieved a lower PDP than the GPU.
This result suggests that IMAX, as a general-purpose accelerator, has the potential to achieve high energy efficiency under specific conditions.
These findings indicate that while the current FPGA prototype and limited offload ratio constrain overall performance, IMAX possesses the potential to become a power-efficient platform for image generation. 
This potential can be realized through the architectural speedups of an ASIC implementation, combined with future improvements in host performance and an increased offload ratio.

%% file: disscution.tex
\section{Discussion}
\label{discussion}
The evaluation in the previous section demonstrated IMAX3's potential to achieve high energy efficiency under certain conditions, while revealing challenges in its end-to-end performance. 
In this section, we analyze these results more deeply. 
Specifically, we focus on the performance of the computational kernels offloaded to IMAX and on the internal behavior of the IMAX system.

\subsection{Kernel Execution Time Analysis by Thread Count}
To clarify the intrinsic computational performance of the IMAX core, which can be obscured in end-to-end measurements, this section analyzes the execution time of only the offloaded quantized dot-product kernels. 
Fig.~\ref{fig:compare_kernel_q3k} and Fig.~\ref{fig:compare_kernel_q80} show the kernel execution time as we vary the number of active IMAX lanes from one to eight.
First, observing the single-lane (1 thread) performance, we see that the FPGA-based IMAX, even operating at just 145\,MHz, achieves faster kernel execution than the host ARM CPU. 
Furthermore, the projected 840\,MHz ASIC version of IMAX demonstrates performance competitive with a high-performance Xeon CPU, confirming the high computational potential of the IMAX core. 
However, a significant performance gap remains when compared to the GPU, which is highly optimized for parallel computation. This difference reflects fundamental disparities in architectural design and the sheer volume of hardware resources.

Next, focusing on scalability, we observe that IMAX's performance improves efficiently up to two lanes but shows diminishing returns at three or more lanes. 
We attribute this performance saturation to the dual-core host CPU in our experimental setup. Although each IMAX lane operates independently, the host CPU manages its data supply and execution control. 
When the number of active lanes exceeds the number of physical host CPU cores, the host's processing capability becomes a bottleneck, preventing the system from fully leveraging the IMAX core's potential. 
This result suggests that to maximize performance in a heterogeneous accelerator system, it is crucial to balance the parallelism of the accelerator with the computational capacity, particularly the core count, of the host system.

\subsection{Breakdown Analysis of IMAX Processing Time}

In this subsection, we investigate the scalability challenges more deeply by analyzing the breakdown of IMAX's processing time, as measured on the FPGA. 
Fig.~\ref{fig:microbenchmark_time} details the time consumed by each processing component during the execution of the Q3\_K and Q8\_0 kernels. 
These components consist of EXEC~(pure computation time on the IMAX computational core), LOAD~(data transfer from main memory to LMM), DRAIN~(result write-back from LMM to main memory), and CONF/REGV/RANGE~(IMAX configuration).



%% file: conclusion.tex
\section{Conclusion}
\label{conclusions}
In this work, we aimed to execute AI image generation on the CGRA accelerator IMAX3 to clarify its performance potential and architectural challenges. 
To achieve this objective, we implemented the primary computational kernels of Stable Diffusion and evaluated their performance on both a current FPGA prototype and a projected ASIC implementation. 
Our evaluation revealed that IMAX3, as a general-purpose platform, possesses high-performance potential. 
Specifically, projections for an ASIC implementation indicate that IMAX3 can achieve kernel-level execution performance competitively with a high-performance CPU and an energy efficiency that can surpass a GPU.

Future work will focus on two key areas. The first is to implement a wider variety of kernels to increase the offload ratio, and the second is to strengthen the integration with a multi-core host to maximize system performance. 
Furthermore, while this study used a standard 512$\times$512 image size for benchmarking, evaluating scalability with larger image resolutions remains an important avenue for future investigation. 
Through these improvements, we will not only mature IMAX3 as a versatile platform but also pave the way for the design of next-generation, specialized AI accelerators.

%% file: main.bbl
\begin{thebibliography}{10}
\providecommand{\url}[1]{#1}
\csname url@samestyle\endcsname
\providecommand{\newblock}{\relax}
\providecommand{\bibinfo}[2]{#2}
\providecommand{\BIBentrySTDinterwordspacing}{\spaceskip=0pt\relax}
\providecommand{\BIBentryALTinterwordstretchfactor}{4}
\providecommand{\BIBentryALTinterwordspacing}{\spaceskip=\fontdimen2\font plus
\BIBentryALTinterwordstretchfactor\fontdimen3\font minus
  \fontdimen4\font\relax}
\providecommand{\BIBforeignlanguage}[2]{{%
\expandafter\ifx\csname l@#1\endcsname\relax
\typeout{** WARNING: IEEEtran.bst: No hyphenation pattern has been}%
\typeout{** loaded for the language `#1'. Using the pattern for}%
\typeout{** the default language instead.}%
\else
\language=\csname l@#1\endcsname
\fi
#2}}
\providecommand{\BIBdecl}{\relax}
\BIBdecl

\bibitem{Xiao_2025_CVPR}
S.~Xiao, Y.~Wang, J.~Zhou, H.~Yuan, X.~Xing, R.~Yan, C.~Li, S.~Wang, T.~Huang,
  and Z.~Liu, ``{OmniGen}: Unified image generation,'' in \emph{Proceedings of
  the Computer Vision and Pattern Recognition Conference (CVPR)}, June 2025,
  pp. 13\,294--13\,304.

\bibitem{Hu_2024_CVPR}
H.~Hu, K.~C. Chan, Y.-C. Su, W.~Chen, Y.~Li, K.~Sohn, Y.~Zhao, X.~Ben, B.~Gong,
  W.~Cohen, M.-W. Chang, and X.~Jia, ``{Instruct-Imagen}: Image generation with
  multi-modal instruction,'' in \emph{Proceedings of the IEEE/CVF Conference on
  Computer Vision and Pattern Recognition (CVPR)}, June 2024, pp. 4754--4763.

\bibitem{llm4gen}
M.~Liu, Y.~Ma, Z.~Yang, J.~Dan, Y.~Yu, Z.~Zhao, Z.~Hu, B.~Liu, and C.~Fan,
  ``{LLM4GEN}: Leveraging semantic representation of {LLMs} for text-to-image
  generation,'' \emph{Proceedings of the AAAI Conference on Artificial
  Intelligence}, vol.~39, pp. 5523--5531, 04 2025.

\bibitem{diffit}
A.~Hatamizadeh, J.~Song, G.~Liu, J.~Kautz, and A.~Vahdat, ``Diffit: Diffusion
  vision transformers for image generation,'' in \emph{Computer Vision -- ECCV
  2024}, A.~Leonardis, E.~Ricci, S.~Roth, O.~Russakovsky, T.~Sattler, and
  G.~Varol, Eds.\hskip 1em plus 0.5em minus 0.4em\relax Cham: Springer Nature
  Switzerland, 2025, pp. 37--55.

\bibitem{NEURIPS2023_821655c7}
Z.~Xue, G.~Song, Q.~Guo, B.~Liu, Z.~Zong, Y.~Liu, and P.~Luo, ``{RAPHAEL}:
  Text-to-image generation via large mixture of diffusion paths,'' in
  \emph{Advances in Neural Information Processing Systems}, A.~Oh, T.~Naumann,
  A.~Globerson, K.~Saenko, M.~Hardt, and S.~Levine, Eds., vol.~36.\hskip 1em
  plus 0.5em minus 0.4em\relax Curran Associates, Inc., 2023, pp.
  41\,693--41\,706.

\bibitem{shehabi2024united}
A.~Shehabi, S.~Smith, A.~Hubbard, A.~Newkirk, N.~Lei, M.~A.~B. Siddik
  \emph{et~al.}, ``2024 united states data center energy usage report,''
  \url{https://escholarship.org/uc/item/32d6m0d1}, Lawrence Berkeley National
  Laboratory, Technical Report LBNL-2001637, 2024, accessed on May 30, 2025.

\bibitem{imax_access}
T.~Akabe, V.~Trung Duong~LE, and Y.~Nakashima, ``{IMAX}: A power-efficient
  multilevel pipelined {CGLA} and applications,'' \emph{IEEE Access}, vol.~13,
  pp. 31\,899--31\,911, 2025.

\bibitem{unetimax}
D.~Thi~Sang, R.~Imamura, T.~Akabe, and Y.~Nakashima, ``{Energy Consumption
  Optimization of Multi-Dimensional U-Nets on CGLA},'' \emph{IEEE Access},
  vol.~13, pp. 29\,476--29\,492, 2025.

\bibitem{imaxcnn2}
R.~Imamura, Z.~Guangxian, S.~D. Thi, H.~L. Pham, R.~Zhang, and Y.~Nakashima,
  ``Energy-efficient 3d convolution using interposed memory accelerator
  extension 2 for^^c2^^a0medical image processing,'' in \emph{Proceedings of
  2023 International Conference on Medical Imaging and Computer-Aided Diagnosis
  (MICAD 2023)}, R.~Su, Y.-D. Zhang, and A.~F. Frangi, Eds.\hskip 1em plus
  0.5em minus 0.4em\relax Singapore: Springer Nature Singapore, 2024, pp.
  62--71.

\bibitem{imaxcnn3}
M.~Tanomoto, S.~Takamaeda-Yamazaki, J.~Yao, and Y.~Nakashima, ``{A CGRA-Based
  Approach for Accelerating Convolutional Neural Networks},'' in \emph{2015
  IEEE 9th International Symposium on Embedded Multicore/Many-core
  Systems-on-Chip}, 2015, pp. 73--80.

\bibitem{uetanicgra}
H.~Uetani and Y.~Nakashima, ``{ Implementation and Evaluation of {LLM} on a
  {CGLA} },'' in \emph{2024 Twelfth International Symposium on Computing and
  Networking (CANDAR)}.\hskip 1em plus 0.5em minus 0.4em\relax IEEE Computer
  Society, 2024, pp. 252--258.

\bibitem{eto2025implementation}
Y.~Eto and Y.~Nakashima, ``Implementation and performance analysis of {LLaMA}
  on a {CGLA},'' in \emph{International Conference on Intelligent Systems and
  Networks (ICISN)}, 2025.

\bibitem{fpgacnn}
C.~Zhang, P.~Li, G.~Sun, Y.~Guan, B.~Xiao, and J.~Cong, ``Optimizing fpga-based
  accelerator design for deep convolutional neural networks,'' in
  \emph{Proceedings of the 2015 ACM/SIGDA International Symposium on
  Field-Programmable Gate Arrays}, ser. FPGA '15.\hskip 1em plus 0.5em minus
  0.4em\relax New York, NY, USA: Association for Computing Machinery, 2015, pp.
  161--170.

\bibitem{vitisai}
J.~Wang and S.~Gu, ``{FPGA Implementation of Object Detection Accelerator Based
  on Vitis-AI},'' in \emph{2021 11th International Conference on Information
  Science and Technology (ICIST)}, 2021, pp. 571--577.

\bibitem{multi_task}
Y.~Lu, X.~Zhai, S.~Saha, S.~Ehsan, and K.~D. McDonald-Maier, ``{FPGA} based
  adaptive hardware acceleration for multiple deep learning tasks,'' in
  \emph{2021 IEEE 14th International Symposium on Embedded Multicore/Many-core
  Systems-on-Chip (MCSoC)}, 2021, pp. 204--209.

\bibitem{cgrasurvey}
A.~Podobas, K.~Sano, and S.~Matsuoka, ``A survey on coarse-grained
  reconfigurable architectures from a performance perspective,'' \emph{IEEE
  Access}, vol.~8, pp. 146\,719--146\,743, 2020.

\bibitem{cgra_cnn2}
J.~Lee and J.~Lee, ``Specializing cgras for light-weight convolutional neural
  networks,'' \emph{IEEE Transactions on Computer-Aided Design of Integrated
  Circuits and Systems}, vol.~41, no.~10, pp. 3387--3399, 2022.

\bibitem{hmap}
D.~Wijerathne, Z.~Li, A.~Pathania, T.~Mitra, and L.~Thiele, ``Himap: Fast and
  scalable high-quality mapping on cgra via hierarchical abstraction,''
  \emph{IEEE Transactions on Computer-Aided Design of Integrated Circuits and
  Systems}, vol.~41, no.~10, pp. 3290--3303, 2022.

\bibitem{cgra_crypto}
S.~D. Thi, H.~Luan~Pham, V.~T. Duong~Le, T.~D. Tran, R.~Imamura, Q.~D.
  Nam~Nguyen, T.~H. Tran, and Y.~Nakashima, ``{Universal 32/64-bit CGRA for
  Lightweight Cryptography in Securing IoT Data Transmission},'' in \emph{2023
  IEEE 16th International Symposium on Embedded Multicore/Many-core
  Systems-on-Chip (MCSoC)}, 2023, pp. 419--425.

\bibitem{llm4fpga}
H.~Chen, J.~Zhang, Y.~Du, S.~Xiang, Z.~Yue, N.~Zhang, Y.~Cai, and Z.~Zhang,
  ``Understanding the potential of {FPGA}-based spatial acceleration for large
  language model inference,'' \emph{ACM Trans. Reconfigurable Technol. Syst.},
  vol.~18, no.~1, Dec. 2024.

\bibitem{llamaf}
H.~Xu, Y.~Li, and S.~Ji, ``{LlamaF}: An efficient {Llama2} architecture
  accelerator on embedded {FPGAs},'' in \emph{2024 IEEE 10th World Forum on
  Internet of Things (WF-IoT)}, 2024, pp. 1--7.

\bibitem{zeng2024flightllm}
S.~Zeng, J.~Liu, G.~Dai, X.~Yang, T.~Fu, H.~Wang, W.~Ma, H.~Sun, S.~Li,
  Z.~Huang, Y.~Dai, J.~Li, Z.~Wang, R.~Zhang, K.~Wen, X.~Ning, and Y.~Wang,
  ``{FlightLLM}: Efficient large language model inference with a complete
  mapping flow on {FPGAs},'' in \emph{Proceedings of the 2024 ACM/SIGDA
  International Symposium on Field Programmable Gate Arrays}, ser. FPGA
  '24.\hskip 1em plus 0.5em minus 0.4em\relax New York, NY, USA: Association
  for Computing Machinery, 2024, pp. 223--234.

\bibitem{llama2fpga}
H.~Xu, X.~Wang, and S.~Ji, ``Towards energy-efficient {Llama2} architecture on
  embedded {FPGAs},'' in \emph{Proceedings of the 33rd ACM International
  Conference on Information and Knowledge Management}, ser. CIKM '24.\hskip 1em
  plus 0.5em minus 0.4em\relax New York, NY, USA: Association for Computing
  Machinery, 2024, pp. 5570--5571.

\bibitem{sdacc}
H.~Zhou, Y.~Liu, H.~Wang, E.~Tang, S.~Li, and Y.~Zhang, ``{SDAcc}: A stable
  diffusion accelerator on {FPGA} via software-hardware co-design,'' in
  \emph{2024 IEEE 32nd Annual International Symposium on Field-Programmable
  Custom Computing Machines (FCCM)}, 2024, pp. 214--214.

\bibitem{sda}
G.~Yang, Y.~Xie, Z.~J. Xue, S.-E. Chang, Y.~Li, P.~Dong, J.~Lei, W.~Xie,
  Y.~Wang, X.~Lin, and Z.~Fang, ``{SDA}: Low-bit {Stable Diffusion}
  acceleration on edge {FPGAs},'' in \emph{2024 34th International Conference
  on Field-Programmable Logic and Applications (FPL)}, 2024, pp. 264--273.

\bibitem{cgra1}
C.~Torng, P.~Pan, Y.~Ou, C.~Tan, and C.~Batten, ``{Ultra-Elastic CGRAs for
  Irregular Loop Specialization},'' in \emph{2021 IEEE International Symposium
  on High-Performance Computer Architecture (HPCA)}, 2021, pp. 412--425.

\bibitem{rombach2022highresolutionimagesynthesislatent}
R.~Rombach, A.~Blattmann, D.~Lorenz, P.~Esser, and B.~Ommer, ``{High-Resolution
  Image Synthesis with Latent Diffusion Models},'' 2022, arXiv:2112.10752.

\bibitem{stable-diffusion.cpp}
Y.-J. Lee, ``{stable-diffusion.cpp}: Inference of {Stable Diffusion} in pure
  {C/C++},'' \url{https://github.com/leejet/stable-diffusion.cpp}, 2022,
  accessed on May 30, 2025.

\bibitem{GerganovLlamaCpp2023}
G.~Gerganov, ``{GitHub-ggerganov/llama.cpp}: {LLM} inference in {C/C++},''
  \url{https://github.com/ggerganov/llama.cpp}, 2023, accessed on May 11, 2025.

\bibitem{Humrick_CortexA72_2016}
M.~Humrick, ``{ARM Cortex-A72 Architecture Deep Dive},''
  \url{https://www.tomshardware.com/reviews/arm-cortex-a72-architecture,4424.html},
  January 2016, accessed on May 30, 2025.

\bibitem{Intel_Xeon_w5-2465X_Specs}
{Intel}, ``{Intel{\textregistered} Xeon{\textregistered} w5-2465X Processor
  (33.75M Cache, 3.10 GHz) Specifications},''
  \url{https://www.intel.co.jp/content/www/jp/ja/products/sku/233415/intel-xeon-w52465x-processor-33-75m-cache-3-10-ghz/specifications.html},
  accessed on May 30, 2025.

\bibitem{NVIDIA_Turing_2018}
Nvidia, ``{NVIDIA Turing GPU Architecture},'' \emph{White Paper
  WP-09183-001\_v01}, vol.~1, 2018.

\bibitem{sauer2023adversarialdiffusiondistillation}
A.~Sauer, D.~Lorenz, A.~Blattmann, and R.~Rombach, ``{Adversarial Diffusion
  Distillation},'' 2023, arXiv:2311.17042.

\bibitem{SynopsysNDDCUltra}
{Synopsys, Inc.}, ``{DC Ultra: Concurrent Timing, Area, Power and Test
  Optimization},''
  \url{https://www.synopsys.com/implementation-and-signoff/rtl-synthesis-test/dc-ultra.html},
  accessed on May 25, 2025.

\end{thebibliography}
